\def\lesssim{\mathrel{\hbox{\rlap{\hbox{\lower4pt\hbox{$\sim$}}}\hbox{$<$}}}}
\def\gtrsim{\mathrel{\hbox{\rlap{\hbox{\lower4pt\hbox{$\sim$}}}\hbox{$>$}}}}
\def\msun{\mbox{$M_{\odot}$}}
\def\teff{$T_{\rm eff}$}

\def\ll_lsun{log$({L/\rm L_{\odot}})$~}
\def\masa_msun{$M/ \rm M_{\odot}$~}
\def\m_mstar{$M/M_{*}$~}

\documentclass{aa}
\usepackage{graphicx}

\newcommand{\bvfreq}{Brunt-V\"ais\"al\"a frequency}
        
\begin{document}

\title{New evolutionary models for massive ZZ Ceti stars. II. The 
effects of crystallization on their pulsational properties}

\author{A. H. C\'orsico$^{1,2}$\thanks{Member of the Carrera del Investigador
Cient\'{\i}fico     y     Tecnol\'ogico,     CONICET,     Argentina.},
L. G. Althaus$^{1,2,3}$\thanks{Member  of the Carrera del Investigador
Cient\'{\i}fico     y     Tecnol\'ogico,     CONICET,     Argentina.},
M.    H.    Montgomery$^4$,    E.   Garc\'{\i}a--Berro$^{3,5}$    \and
J. Isern$^{3,6}$}

\offprints{A. H. C\'orsico}

\institute{$^1$Facultad  de  Ciencias Astron\'omicas  y Geof\'{\i}sicas, 
Universidad  Nacional de  La Plata, Paseo  del  Bosque s/n,  (1900)  
La  Plata,  Argentina.\\ 
$^2$Instituto  de Astrof\'{\i}sica  La Plata, IALP, CONICET\\
$^3$Departament   de  F\'\i   sica   Aplicada,  Universitat
           Polit\`ecnica de Catalunya, Escola Polit\`ecnica Superior
           de Castelldefels, Av. del Canal Ol\'\i mpic, s/n,
           08860 Castelldefels, Spain\\
$^4$Department of Astronomy, University of Texas, Austin, TX 78712, USA\\
$^5$Institut d'Estudis Espacials  de Catalunya, Ed.  Nexus,
           c/Gran Capit\`a 2, 08034 Barcelona, Spain.\\  
$^6$Institut de Ci\`encies de l'Espai (CSIC)\\
\email{acorsico@fcaglp.unlp.edu.ar, leandro@fa.upc.es, mikemon@ast.cam.ac.uk, 
garcia@fa.upc.es,\\ isern@ieec.fcr.es}}
\date{Received; accepted}


\abstract{In view of  recent claims that asteroseismology could supply
invaluable insights into the  crystallization process occurring in the
interiors of  massive white dwarf stars,  we present in  this work new
pulsational  calculations for  improved carbon-oxygen  DA  white dwarf
models suitable for the study of massive ZZ Ceti stars. The background
models employed in  this study, presented in detail  in a recent paper
by Althaus et al.  (2003), are the result of the complete evolution of
massive  white  dwarf  progenitors  from the  zero-age  main  sequence
through the Asymptotic Giant Branch  (AGB) and mass loss phases to the
white dwarf regime.  Abundance changes are accounted for by means of a
full coupling between nuclear  evolution and time-dependent mixing due
to convection,  salt fingers,  and diffusive overshoot.   In addition,
time-dependent  element diffusion  for multicomponent  gases  has been
considered  during  the white  dwarf  evolution.  Crystallization  and
chemical rehomogenization due to phase separation upon crystallization
in the core of our models  have been fully considered.  The effects of
crystallization on  the period spectrum  of these massive  white dwarf
models are  assessed by  means of a  detailed pulsational  analysis of
linear, nonradial,  adiabatic gravity modes.  To  properly account for
the effects of  the presence of a solid phase in  the models we employ
special   conditions  on   the  oscillation   eigenfunctions   at  the
solid-liquid  interface.   We  find  that  the  theoretical  pulsation
spectrum is  strongly modified when crystallization  is considered, in
particular concerning the mode  trapping properties of the equilibrium
models. We show that the strong  mode trapping seen in the models with
overshooting can be reproduced by  means of a simple analytical model.
We  also discuss  at some  length the  implications of  our  study for
BPM~37093,  the  most  massive  ZZ  Ceti  star  presently  known.   In
particular, we attempt to  place constraints on the physical processes
occurring prior to the formation of this white dwarf.  We find that if
BPM~37093 has  a stellar mass  of $\approx 1.00$ \msun\,  its observed
spectrum  could bear the  signature of  overshoot episodes  during the
helium core burning.
}

\authorrunning{C\'orsico et al.}

\titlerunning{The effects of crystallization on the pulsational properties}

\maketitle


\section{Introduction}

ZZ Ceti (or  DAV) stars are cool, hydrogen-rich  pulsating white dwarf
stars belonging  currently to the  most extensively studied  family of
degenerate  pulsators ---  see Gautschy  \&  Saio (1995,  1996) for  a
review.  Their  pulsating nature  is evident from  periodic brightness
variations\footnote{Also fluctuations in radial velocity have recently
been measured in  some ZZ Ceti stars, see van  Kerkwijk et al.  (2000)
and  Kotak   et  al.   (2002).}   caused   by  spheroidal,  non-radial
$g$(gravity)-modes  of  low degree  ($\ell  \leq  2$).  Excitation  of
100--1200 s oscillation modes is likely to be due to the action of the
so  called   {\sl  convective  driving}   mechanism  (Brickhill  1991,
Goldreich \& Wu  1999), although the early works  of Dolez \& Vauclair
(1981) and Winget  et al.  (1982) postulated  the $\kappa-\gamma$
mechanism as responsible  for driving.  ZZ Ceti stars  are well known
to pulsate  in a ``pure'' instability strip  at effective temperatures
(\teff) between about 11000 and  12400 K.  Relevant studies devoted to
exploring the  pulsational properties  of ZZ Ceti  stars are  those of
Brassard  et al.  (1991b;  1992ab), Gautschy  et al.   (1996), Bradley
(1996, 1998a, 2001) and C\'orsico et al. (2001, 2002) amongst others.

Over the last years, pulsation studies of ZZ Ceti stars --- as well as
of DBV,  DOV and PNNV stars,  the other categories  of pulsating white
dwarfs ---  through asteroseismology have become  a valuable technique
for  sounding  the  white  dwarf  interiors  and  evolution.   Indeed,
asteroseismological  inferences  have  provided  independent  valuable
constraints to fundamental quantities  such as core composition, outer
layer chemical stratification and  stellar mass (Pfeiffer et al. 1996;
Bradley 1998b, 2001).

\begin{figure}
\centering
\includegraphics[clip,width=250pt]{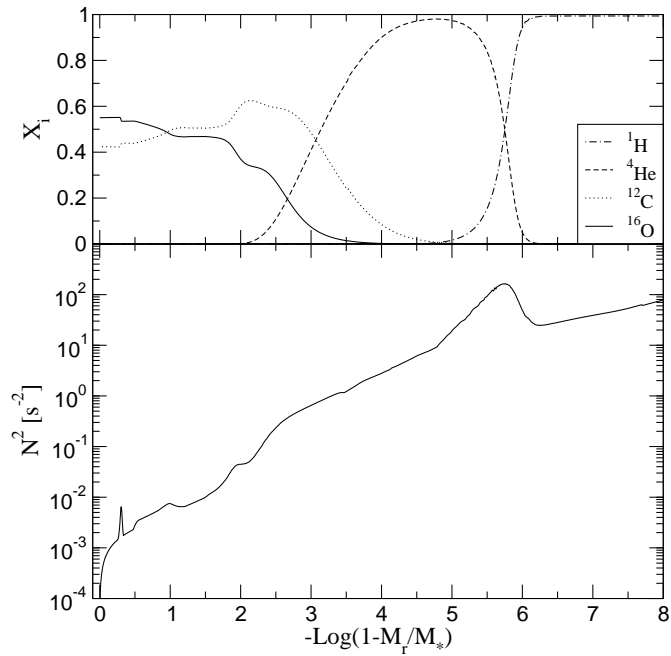}
\caption{Abundance by mass of $^{1}$H, $^{4}$He, $^{12}$C and $^{16}$O 
(upper panel) and the  run of the Brunt-V\"ais\"al\"a frequency (lower
panels)  as  a  function  of  the  outer mass  fraction  $q$  for  the
0.936-$M_{\odot}$  white dwarf  remnant  of 7.5-$M_{\odot}$  evolution
(sequence  NOV) near  the ZZ  Ceti  instability strip.   The model  is
characterized  by  $\log(L/L_\odot)=-2.92$  and  $\log T_{\rm  eff}  =
4.06$.}
\label{xbvnov}
\end{figure}

\begin{figure}
\centering
\includegraphics[clip,width=250pt]{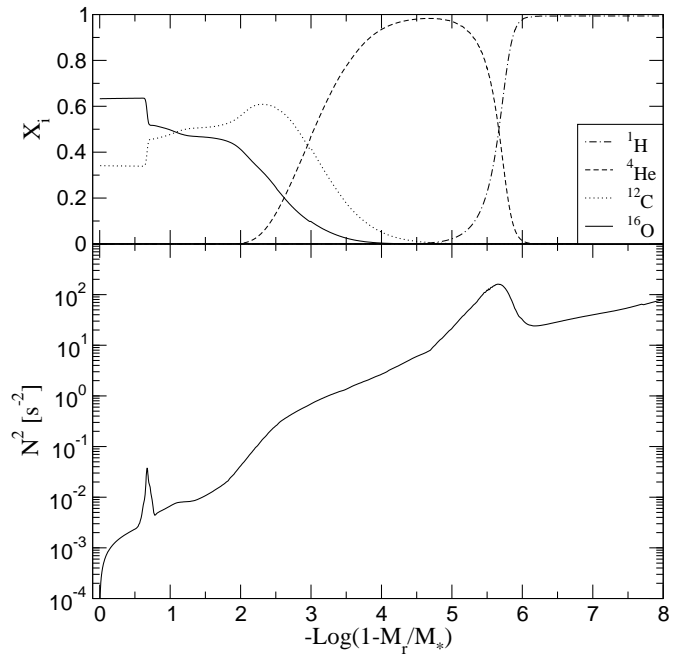}
\caption{Same as Fig. ~\ref{xbvnov} but for the 0.94-$M_{\odot}$ white 
dwarf  remnant   of  6-$M_{\odot}$  evolution   (sequence  OV).   Core
overshooting leads  to a inner  chemical profile quite  different from
the situation  in which  no extra mixing  beyond the  fully convective
core  is considered  (Fig.   ~\ref{xbvnov}).  In  particular note  the
almost-discontinuity in the chemical profile at $\log (1-M_r/M_*)
\approx -0.7$.}
\label{xbvov}
\end{figure}

An outstanding  application of  white dwarf asteroseismology  that has
drawn the  attention of researchers  is related to the  possibility of
placing  observational constraints on  the crystallization  process in
the very dense interiors of white dwarfs.  Crystallization in the core
of white  dwarfs was predicted  theoretically about 40 years  ago (Van
Horn 1968; see also Kirzhnits 1960; Abrikosov 1960 and Salpeter 1961),
but it was not until recently  that researchers have been able to peer
into  the  crystallized structure  of  white  dwarfs.   This has  been
prompted by the discovery of  pulsations in the star BPM~37093 (Kanaan
et  al.  1992),  a  massive  ZZ  Ceti star  which  should  be  largely
crystallized (Winget el al.   1997). In a very detailed investigation,
Montgomery   \&   Winget  (1999)   have   explored   the  effects   of
crystallization on  the period pattern of massive  white dwarf stellar
models.  Amongst  others findings,  these authors concluded  that some
periods are notoriously sensitive  to changes in the crystallized mass
fraction of their models.   They concluded that pulsating white dwarfs
are, in principle, very promising  objects to place constraints on the
crystallization processes in stars.

Very  recently,  Metcalfe,  Montgomery, \&  Kanaan  (2004)  have
performed  asteroseismological fits  to the  pulsation periods  of the
star BPM 37093.   In their models, the crystallized  mass fraction was
treated  as a  free  parameter, with  the  goal of  \emph{empirically}
determining the degree of crystallization for a given stellar mass and
core composition. Their  preliminary results are that BPM  37093 has a
crystallized mass fraction greater  than 50\%, with values possibly as
high as 90\%,  although they indicated that an  exploration of a finer
grid of models  (in both stellar mass and  crystallized mass fraction)
will be necessary in order to make conclusive statements.

Crystallization  has consequences  for the  carbon/oxygen distribution
within  the  core of  a  white  dwarf.  As  a  matter  of fact,  solid
theoretical evidence  suggests that, when  crystallization occurs, the
oxygen content in the solid phase  is enhanced relative to that in the
original fluid  phase (Stevenson 1980; Garc\'{\i}a-Berro  et al. 1988;
Ichimaru et al.  1988; Segretain  \& Chabrier 1993).  As a result, the
carbon content  in the fluid  surrounding the solid core  is enhanced.
Since  carbon   is  lighter  than  oxygen,  these   fluid  layers  are
Rayleigh-Taylor  unstable,  and  the  ensuing convective  mixing  will
redistribute  the abundances  and lead  to flat  profiles in  a region
whose  size depends  on the  initial  composition profile  and on  the
degree  of chemical  enhancement produced  during  the crystallization
process (Isern  et al. 1997; Salaris  et al.  1997;  Montgomery et al.
1999).  These  mixing episodes could  be particularly relevant  in the
context of  carbon/oxygen cores  with varying chemical  profiles since
they,  in   principle,  could  be   smoothed  out  by   such  chemical
rehomogenization.  To  the best of our  knowledge, pulsational studies
of   massive   white  dwarfs   taking   into   account  the   chemical
rehomogenization induced by phase  separation have not been performed;
this  effect was  not considered  in the  pulsational  calculations of
Montgomery \& Winget (1999), although  it was fully taken into account
in the  calculations of  phase separation and  cooling delay  in white
dwarfs  by Salaris  et  al.  (1997), Montgomery  et  al.  (1999),  and
Salaris et  al. (2000).  In view  of these considerations,  one of the
main motivations  for the present paper concerns  the full exploration
of the pulsational properties of  massive white dwarfs that takes into
account the effects of phase separation in carbon/oxygen cores.

\begin{figure}
\centering
\includegraphics[clip,width=250pt]{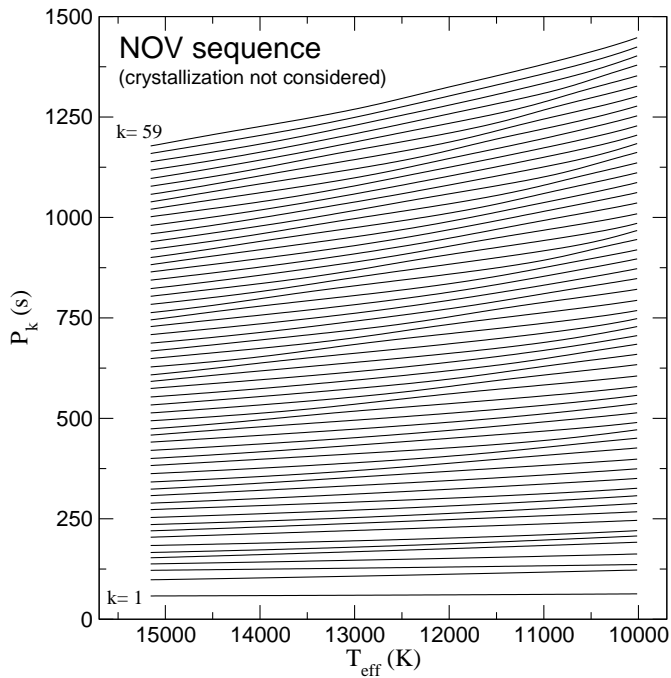}
\caption{Evolution  of the  periods for $\ell= 2$  modes as a function  
of   the    effective   temperature,   corresponding    to   the   NOV
model-sequence. The effects of  crystallization have been neglected in
computing the eigenmodes.}
\label{pncnrnov}
\end{figure}
  
\begin{figure}
\centering
\includegraphics[clip,width=250pt]{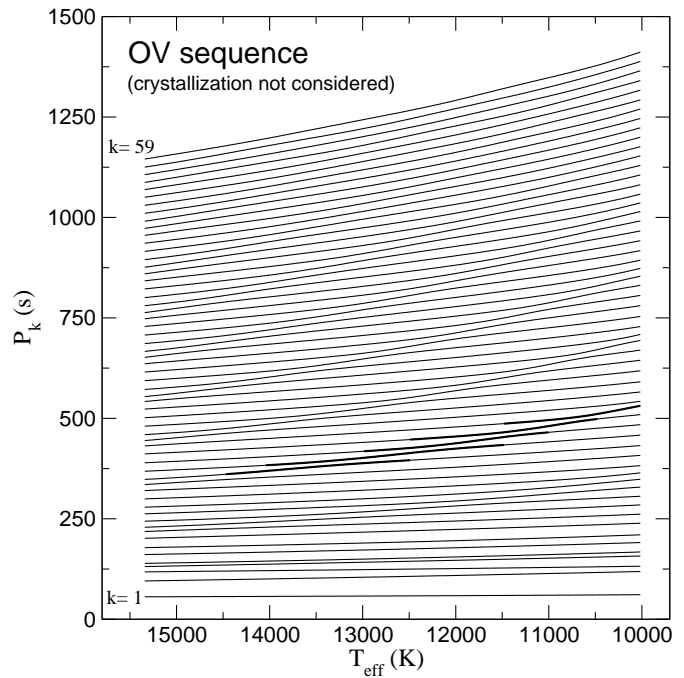}
\caption{Same as Fig. ~\ref{pncnrnov}, but  for the OV model-sequence.  
Thick   lines   emphasize   some   examples  of   modes   experiencing
bumping/avoided crossing phenomena.}
\label{pncnrov}
\end{figure}

Recently,  Althaus et  al. (2003)  ---  hereinafter Paper  I ---  have
presented  new stellar  models for  massive  ZZ Ceti  stars, based  on
detailed  evolutionary  calculations  that  account for  the  complete
evolution  of   the  white  dwarf  progenitor.   In  particular,  such
calculations include diffusive mechanical overshooting during the core
burning phases of the progenitor  star.  In addition, the evolution of
the  chemical  abundance distribution  due  to time-dependent  element
diffusion during the  whole white dwarf regime has  been considered in
Paper I. In that work we concluded that, as a result of the smoothness
of  the  chemical profile  caused  by  diffusion  processes, the  mode
trapping  due  to outer  chemical  interfaces  is notably  diminished,
irrespective  of the  occurrence of  core overshooting.   Instead, the
theoretical pulsational  spectrum is characterized by  the presence of
pronounced  non-uniformity in  the spacing  of consecutive  periods in
models with core overshooting, at variance with the situation in which
this mixing  process is neglected.   In particular, we found  that the
pulsational spectrum in models  with overshooting are dominated by the
presence of  ``core trapped'' modes, characterized  by relatively high
values  of the  oscillation kinetic  energy and  strong minima  in the
period spacing diagrams.

The  main  conclusion  drawn  in  Paper  I  is  that  the  pulsational
properties  of massive  ZZ Ceti  stars  become very  sensitive to  the
occurrence  of  extra  mixing  episodes  that take  place  beyond  the
formally  convective  core  during  the central  helium  burning,  for
instance core  overshoot. Such  mixing episodes give  rise to  a sharp
variation  of  the core  chemical  composition  that leave  noticeable
signatures  on  the theoretical  period  spectrum  of pulsating  white
dwarfs.  However,  the fact that  the eigenfunctions of  $g$-modes are
expected to have very low amplitudes  in the solid core --- because of
the non-zero  shear modulus  of the solid   (Montgomery  \& Winget
1999) --- as  well as the mixing episodes  induced by crystallization
were  two effects  not addressed  in Paper  I.  Therefore,  an initial
question  to  be  answered  in  this paper  is  whether  the  chemical
discontinuity  caused by  core overshoot  could  be wiped  out by  the
chemical  rehomogenization, and  more  importantly, what  implications
this  would have  for the  pulsational properties  of massive  ZZ Ceti
stars.  In  this connection,  another  aim of  our  work  is to  place
constraints on the  stellar mass and \teff\ values  at which we should
expect a pulsational pattern without any signature of core overshoot.

We  believe that  a re-examination  of the  pulsational  properties of
massive white dwarf stars deserves to  be done in the frame of our new
massive ZZ Ceti model stars.   Specifically, our aim is to explore the
pulsational properties of the white  dwarf models presented in Paper I
taking crystallization  self-consistently into account.   Section \S 2
contains  a brief  description  of  the main  physical  inputs of  the
models. In \S 3 we explore  the response of the pulsational pattern to
the  presence  of  crystallization.    Section  \S  4  is  devoted  to
discussing   the  implications  of   our  results   for  observational
expectations and, finally, in \S 5 we make some concluding remarks.

\section{Input physics and evolutionary sequences}

\begin{figure}
\centering
\includegraphics[clip,width=250pt]{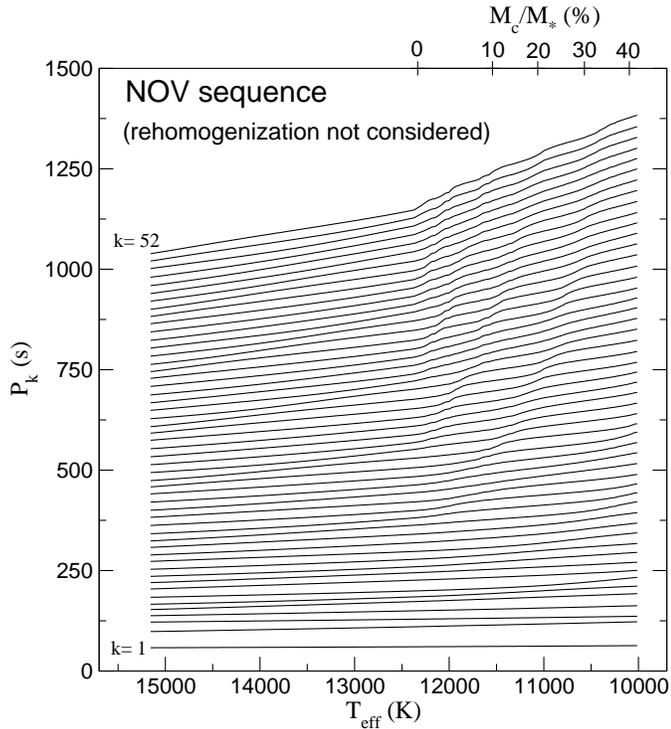}
\caption{Same as Fig. ~\ref{pncnrnov}, but for  the case in which  the 
presence of a solid core has been considered in the computation of the
eigenmodes.   Chemical rehomogenization  upon crystallization  has not
been  taken  into account  in  the  stellar  model.  The  upper  scale
measures the crystallized mass fraction.}
\label{cnrnov}
\end{figure}

\begin{figure}
\centering
\includegraphics[clip,width=250pt]{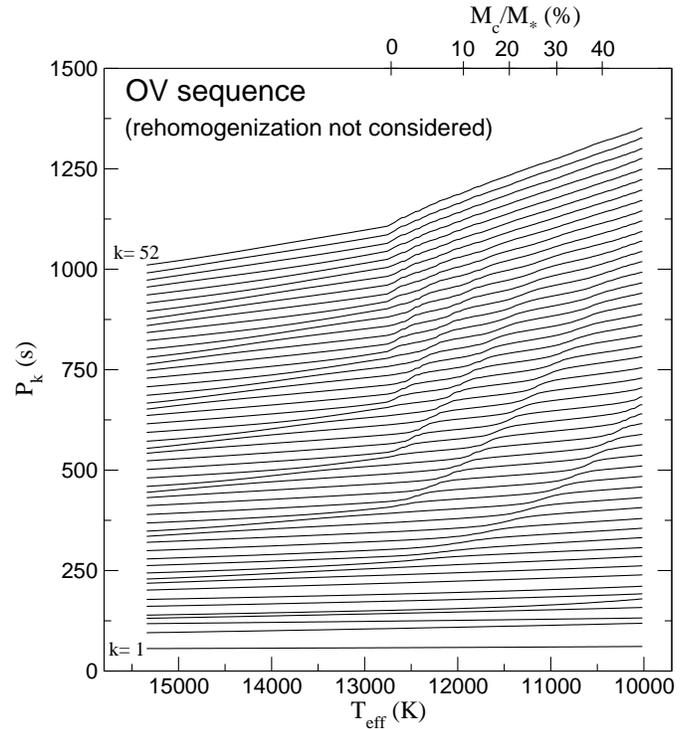}
\caption{Same as Fig. ~\ref{cnrnov}, but for the OV model-sequence.}
\label{cnrov}
\end{figure}

Our massive, carbon/oxygen-core white  dwarf models appropriate for ZZ
Ceti  stars  have been  obtained  with  the  LPCODE evolutionary  code
described at  length in Paper I  and references therein.   The code is
based on an up-to-date and detailed physical description, particularly
regarding the processes responsible for the chemical changes.  In this
section, we  summarize the input physics  of our stellar  code and the
main characteristics of our evolutionary models.

LPCODE  uses  OPAL  radiative  opacities  ---  including  carbon-  and
oxygen-rich compositions  --- for arbitrary  metallicity from Iglesias
\& Rogers  (1996) and molecular  opacities from Alexander  \& Ferguson
(1994).  The equation of  state includes partial ionization, radiation
pressure,  ionic  contributions,  partially degenerate  electrons  and
Coulomb  interactions.  For  the  white dwarf  regime,  we include  an
updated  version of  the  equation  of state  of  Magni \&  Mazzitelli
(1979).  Neutrino emission rates and high-density conductive opacities
are taken from the works of  Itoh and collaborators --- see Althaus et
al.   (2002).  A  total  of  34 thermonuclear  reaction  rates and  16
isotopes characterizes our nuclear network that describes the hydrogen
--- proton-proton chain  and CNO bi-cycle --- and  helium burning, and
carbon ignition.   Nuclear reaction rates  are taken from  Caughlan \&
Fowler (1988) and Angulo et al.  (1999).

As for the abundance changes,  we consider a time-dependent scheme for
the  simultaneous  treatment of  chemical  changes  caused by  nuclear
burning  and  mixing  processes.   Specifically, the  changes  in  the
abundances  for all  chemical elements  are  described by  the set  of
equations

\begin{equation} \label{ec1}
\left( \frac{d \vec{Y}}{dt} \right) = 
\left( \frac{\partial \vec{Y}}{\partial t} \right)_{\rm nuc} +
\frac{\partial}{\partial M_r} \left[ (4\pi r^2 \rho)^2 D 
\frac{\partial \vec{Y}}{\partial M_r}\right],  
\end{equation} 

\noindent with  $\vec{Y}$  being  the  vector  containing  the  number 
fraction  of all  considered  nuclear species.   Here,  mixing due  to
convection,  salt  finger and  overshoot  is  treated  as a  diffusion
process which  is described  by the second  term of  Eq.  (\ref{ec1}).
The efficiency  of convective and  salt-finger mixing is  described by
appropriate diffusion  coefficients, $D$,  which are specified  by the
treatment of  convection.  In  particular, we considered  the extended
mixing  length  theory  of  convection  for  fluids  with  composition
gradients  developed   by  Grossman  et  al.   (1993)   in  its  local
approximation  as given by  Grossman \&  Taam (1996).   This treatment
applies  in convective,  semiconvective\footnote{Semiconvective mixing
--- see Straniero  et al.  (2003) ---  has not been  considered in our
models.}  and salt finger instability  regimes.  The first term of Eq.
(\ref{ec1})   gives  the  abundance   changes  due   to  thermonuclear
reactions, changes which are  fully coupled to mixing processes.  This
term is linearized  following the implicit scheme of  Arnett \& Truran
(1969).  For details concerning the numerical scheme employed to solve
Eq. (1) see Paper I.

\begin{figure}
\centering
\includegraphics[clip,width=250pt]{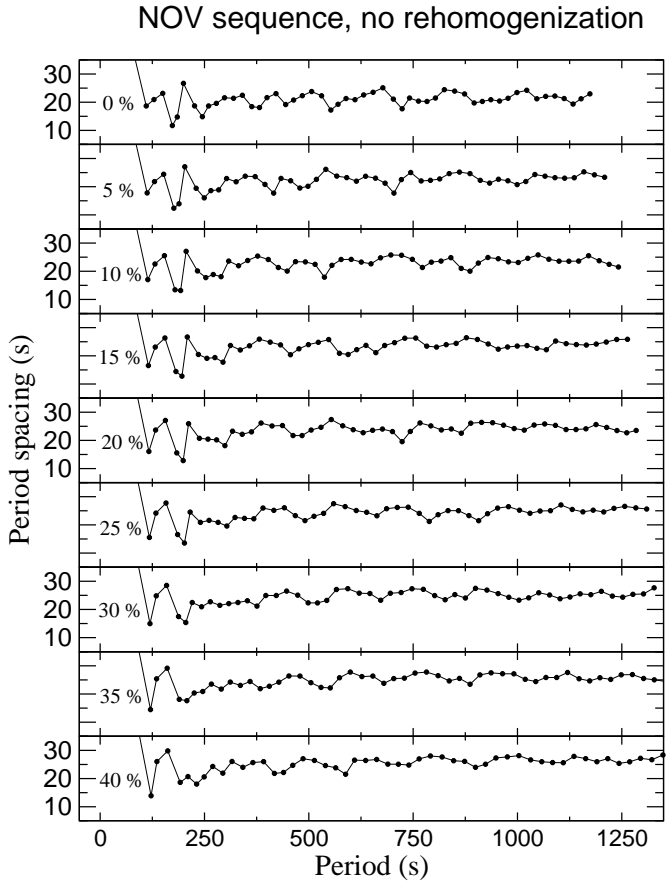}
\caption{The  forward  period  spacing ($\Delta P_k=  P_{k+1}-P_k$) in 
terms  of the periods  for $\ell=  2$ modes  corresponding to  the NOV
model-sequence.   Each  panel  shows  the  predictions  for  different
percentages   of    the   crystallized   mass    fraction.    Chemical
rehomogenization was not considered.}
\label{dpcnrnov}
\end{figure}

\begin{figure}
\centering
\includegraphics[clip,width=245pt]{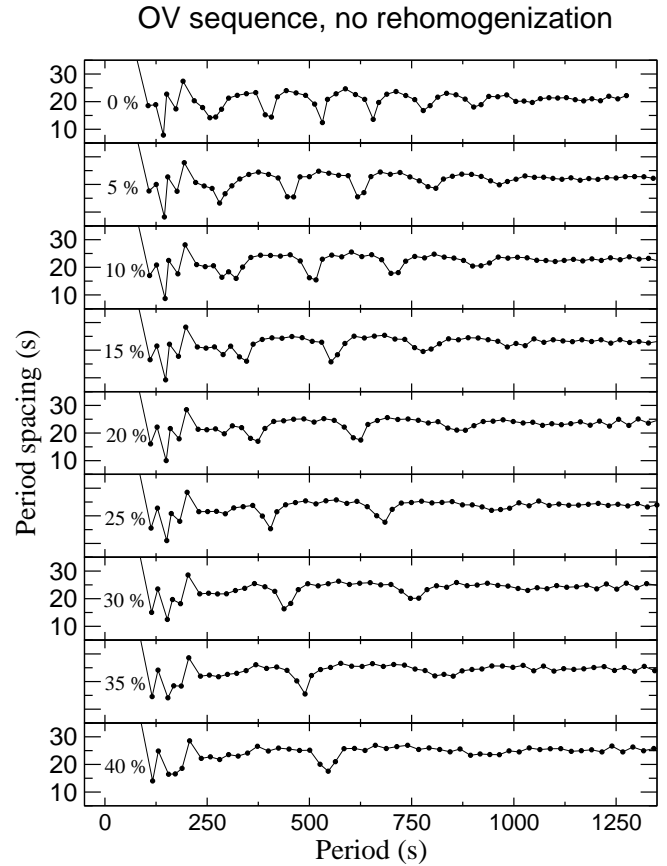}
\caption{Same  as  Fig. ~\ref{dpcnrnov}, but  for the  case  of the OV 
model-sequence.} 
\label{dpcnrov}
\end{figure}

The  evolution  of  the  chemical  abundance  distribution  caused  by
diffusion processes during the whole white dwarf regime has been taken
into  account  in this  work.   Our  time-dependent element  diffusion
treatment, based on the formulation for multicomponent gases presented
by  Burgers  (1969), considers  gravitational  settling, chemical  and
thermal  diffusion\footnote{Radiative levitation has  been neglected.}
for  the nuclear  species $^{1}$H,  $^{3}$He, $^{4}$He,  $^{12}$C, and
$^{16}$O.   In  this  way, we  avoid  the  use  of the  trace  element
approximation usually invoked in most ZZ Ceti studies.

As for overshooting, we  have included time-dependent overshoot mixing
during all  pre-white dwarf  evolutionary stages.  We  have considered
exponentially  decaying diffusive  overshooting above  and  below {\sl
any} formally  convective region, including the  convective core (main
sequence and  central helium burning phases),  the external convective
envelope  and  the  short-lived  helium-flash  convection  zone  which
develops during  the thermal  pulses.  Specifically, we  have followed
the formalism of Herwig (2000).

The effect  of crystallization on  the pulsational pattern of  ZZ Ceti
stars constitutes the central point  of this paper.  For a white dwarf
model characterized by  a given $T_{\rm eff}$ value,  stellar mass and
chemical composition, LPCODE provides a formally self-consistent value
of  the   crystallized  mass  fraction  ($M_{\rm   c}/M_*$).   In  our
calculations, crystallization is assumed to occur when $\Gamma >
180$ (Ogata  \& Ichimaru 1987; Stringfellow, De  Witt \& Slattery
1990), where  $\Gamma \equiv Z^2 e^2  / \overline{r} k_{\rm  B} T$ is
the ion coupling constant.  In order to evaluate self-consistently the
perturbations   caused   by   a   crystal/fluid   interface   on   the
eigenfunctions  we shall  adopt  special boundary  conditions (see  \S
3.1).   In   addition,  the  chemical  redistribution   due  to  phase
separation  has  been  taken  into  account  following  the  procedure
described in Montgomery et al.   (1999) and Salaris et al.  (1997). To
assess the  enhancement of oxygen  in the crystallizing core,  we have
employed the  phase diagram of Segretain \&  Chabrier (1993).  Details
concerning the algorithm  used to compute the mixing  processes due to
crystallization will be described in \S 3.5 below.

\begin{figure}
\centering
\includegraphics[clip,width=250pt]{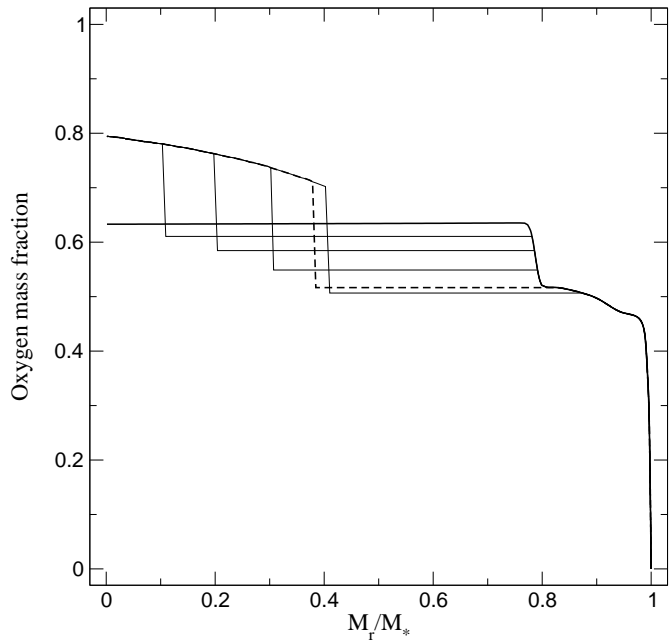}
\caption{Internal $^{16}$O chemical profiles in terms of the fractional 
mass  for the  OV  sequence corresponding  to  various percentages  of
crystallization.   Thick-solid line  corresponds to  the  profile just
before the onset of crystallization,  and thin lines correspond to the
chemical   profiles  resulting   from  chemical   rehomogenization  at
increasing percentages of crystallized mass fraction, from 10 to 40 \%
in steps  of 10 \%. The curve  at 40 \% of  crystallized mass fraction
corresponds  to $T_{\rm  eff}  \approx 10000$  K.   Thick dashed  line
represents the  oxygen profile by the time  the chemical discontinuity
left by  core overshooting has  been completely wiped out  by chemical
rehomogenization  (when  $\approx   37  \%$  of  the  core   is  in  a
crystallized state).}  
\label{rehomo}
\end{figure}

An  important  aspect   of  the  present  study  is   to  explore  the
implications of  the occurrence  of overshooting during  the pre-white
dwarf evolution for the pulsational properties of crystallized ZZ Ceti
stars.  To  this end, we consider the  evolutionary sequences analyzed
in Paper I, that is sequence  NOV based on the complete evolution of a
7.5-\msun\ initial mass star in which overshooting was not considered,
and  sequence OV  based on  the  evolution of  a 6-\msun  \ star  with
overshooting (a solar-like initial composition has been adopted).  For
both sequences, the mass of  the resulting carbon/oxygen core is quite
similar ($\approx  0.94$ \msun).  This  has enabled us to  compare the
oscillation properties of crystallized  ZZ Ceti stars characterized by
the  same  stellar mass  but  being the  result  of  the evolution  of
progenitor stars with different initial masses. It is worth mentioning
that the ZZ Ceti models we  employed in our analysis are the result of
the {\sl  complete} evolution of massive white  dwarf progenitors from
the zero-age main sequence through the thermally pulsing and mass loss
phases  to  the  white  dwarf  regime.   In  particular,  sequence  OV
experiences  the third  dredge-up and  hot bottom  burning  during the
thermally  pulsing phase  on the  asymptotic giant  branch  (AGB). For
details  concerning the  evolutionary properties  of our  sequences we
refer the reader  to Paper I.  In what  follows, we restrict ourselves
to commenting on the main results for the inner chemical composition.

The  upper panels of  Figs.  ~\ref{xbvnov}  and ~\ref{xbvov}  show the
chemical profiles  at the ZZ Ceti  stage for sequences NOV  and OV. As
demonstrated in Paper I, except for  the inner part of the core, where
the  diffusion time scale  becomes much  longer than  the evolutionary
time  scale, element  diffusion  is so  efficient  that the  resulting
abundance distribution  at the  ZZ Ceti stage  does not depend  on the
occurrence  of  overshooting in  the  convective  envelope during  the
thermally pulsing AGB phase.  In fact, the resulting external chemical
profile   at  the   ZZ  Ceti   stage   is  quite   similar  for   both
sequences.  Remarkably  enough,  our  models are  characterized  by  a
chemical   interface   in  which   helium,   carbon   and  oxygen   in
non-negligible abundances coexist, an  interface which, when the white
dwarf reaches the ZZ Ceti  stage, has extended appreciably as a result
of   chemical  diffusion.   Needless   to  say,   the  trace   element
approximation would  not be  in this case  an appropriate  approach to
treat element diffusion.

However, overshoot episodes occurring  during the core burning phases,
particularly  during  the  helium  core  burning,  leave  recognizable
features in the  inner chemical profile of a massive  ZZ Ceti star, as
it is documented  by Fig. ~\ref{xbvov}. That is,  the innermost region
of such  stars keeps a record  of the extra  mixing experienced during
the  pre-white  dwarf  evolution.    In  particular,  note  the  sharp
variation  of  the carbon/oxygen  profile  at  $\log  q \approx  -0.7$
($q=1-M_r/M_*$).  As  shown  in  Paper  I,  this  is  responsible  for
noticeable structure in the period spacing diagrams.  The shape of the
chemical profile  towards the central region of  our OV model-sequence
is typical  for situations in  which additional mixing beyond  what is
predicted  by  Schwarzschild  criterion  for convective  stability  is
allowed.  The  occurrence of  such mixing episodes,  particularly core
overshooting and/or  semiconvection, is suggested  by both theoretical
and  observational  evidence.  In  particular,  extra mixing  episodes
beyond  the external  border of  the fully  convective core  that take
place towards the end of central helium burning have a large influence
on the  carbon and  oxygen distribution in  the core of  white dwarfs.
Recently, Straniero  et al. (2003) have presented  a detailed analysis
of the  inner chemical  abundance in a  3-\msun\ star  model resulting
from different  extra mixing processes  during the late stage  of core
helium burning phase.  In  particular, they conclude that models which
incorporate semiconvection or  a moderate mechanical overshoot applied
to  core and  convective  shells,  predict a  sharp  variation of  the
chemical composition in the  carbon/oxygen core, in agreement with our
results for OV sequence.

\section{Pulsational analysis}

In this  section we shall explore  in detail the  consequences for the
theoretical  period spectrum  of our  white  dwarf models  due to  the
effects  of  crystallization.  In  all  the cases,  we  shall  analyze
model-sequences  NOV and OV  in models  with $T_{\rm  eff}$ decreasing
from  $\approx   15000$  to  $\approx  10000$  K,   which  covers  the
temperature  range for  the observed  ZZ Ceti  instability  strip.  In
showing our results we shall  concentrate on $\ell= 2$ $g$-modes only;
the results adopting other values of $\ell$ are qualitatively similar.

Before discussing  the pulsational results, we  briefly describe below
the pulsational code and the  treatment employed to assess the various
pulsational quantities.

\begin{figure}
\centering
\includegraphics[clip,width=250pt]{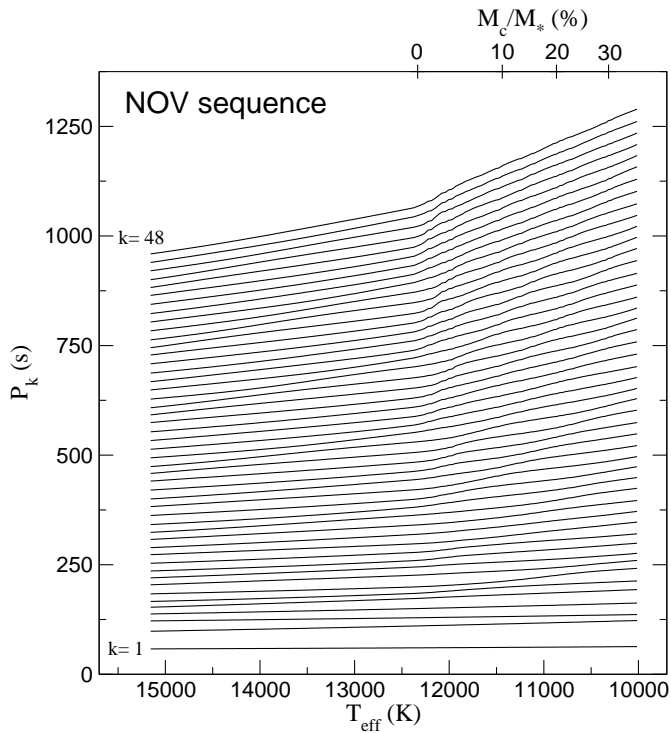}
\caption{Evolution of the periods for $\ell= 2$ modes as a function of  
the  effective temperature, corresponding  to the  NOV model-sequence.
The  full effects  of  crystallization have  been  considered both  in
assessment the  final shape  of the internal  chemical profile  and in
computing the eigenmodes.}
\label{pcrnov}
\end{figure}

\begin{figure}
\centering
\includegraphics[clip,width=250pt]{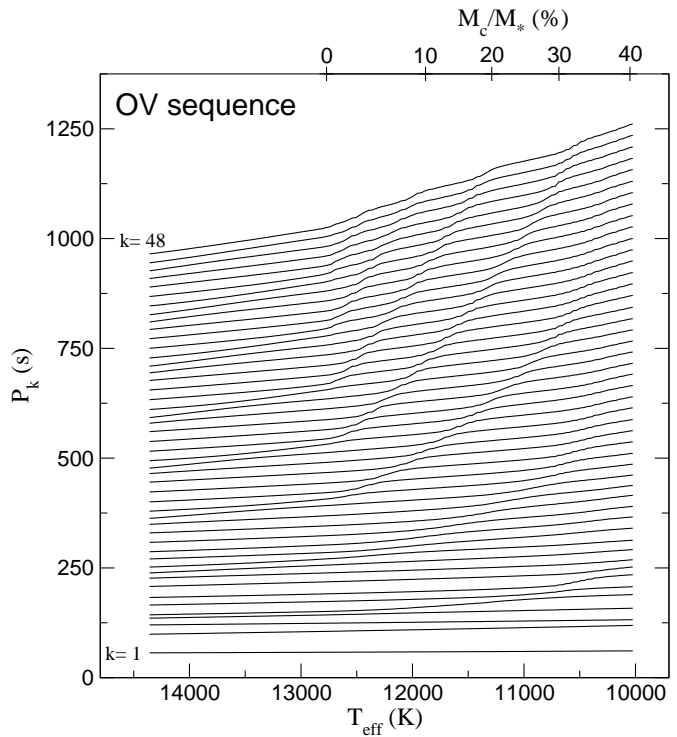}
\caption{Same of Fig. ~\ref{pcrnov}, but for the OV model-sequence} 
\label{pcrov}
\end{figure}

\subsection{Pulsation code}

For computing adiabatic, nonradial $g$-modes  of the ZZ Ceti models we
employ the same pulsational code  as in C\'orsico et al. (2001, 2002),
with   appropriate   modifications  for   handling   the  effects   of
crystallization on oscillation eigenmodes. Briefly, the code, which is
coupled  to the  LPCODE evolutionary  code,  is based  on the  general
Newton-Raphson technique to solve the full set of linearized equations
governing spheroidal\footnote{We have  not considered torsional modes,
since these modes are characterized by very short periods --- up to 20
s; see Montgomery \& Winget (1999) --- and are not observed in ZZ Ceti
stars.},  adiabatic,  nonradial  pulsations of  spherically  symmetric
stars.  The code provides the  eigenperiod $P_k$ (being $k$ the radial
overtone  of mode) and  the dimensionless  eigenfunctions $y_1,\cdots,
y_4$ --- see Unno et al. (1989) for their definition. Useful pulsation
quantities,  such  as  the  oscillation  kinetic  energy,  the  weight
function, and  the variational period  --- see Kawaler et  al.  (1985)
--- are  also  provided by  our  pulsational  code  for each  computed
eigenmode. Finally,  the asymptotic period  spacing is computed  as in
Tassoul et al. (1990).

The boundary conditions are those  given by Osaki \& Hansen (1973) and
the normalization  condition adopted at  the stellar surface  is $y_1=
1$.  As for the inner boundary conditions of our crystallizing models,
we have adopted the ``hard-sphere'' boundary condition, which has been
shown   by   Montgomery  \&   Winget   (1999)  to   be   a   realistic
representation. In fact, the  amplitude of eigenfunctions of $g$-modes
is  drastically reduced below  the solid/liquid  interface due  to the
non-zero shear modulus of the solid, as compared with the amplitude in
the fluid region --- see figure 4 of Montgomery \& Winget (1999).

Specifically, the  hard-sphere boundary condition at  the radial shell
corresponding    to   the    outward-moving    crystallization   front
($r_{\rm c}=r(M_{\rm c})$) reads

\begin{equation}
\begin{tabular}{l}
$y_1= 0,$\\
\\
$y_2=$ arbitrary,\\
\\
$\ell\ y_3-y_4= 0$.\\
\end{tabular}
\end{equation}

\noindent Here,    $y_1$   and    $y_2$   represent  the   radial  and 
tangential displacements,  respectively, and  $y_3$ and $y_4$  are the
Eulerian   perturbation  of  the   gravitational  potential   and  its
derivative.   Note that  the last  condition is  the same  as  for the
normal case  in which the  core is in  a fluid state and  the boundary
condition  is applied  at the  stellar centre  --- see  Appendix  B of
Montgomery \& Winget (1999).

\begin{figure}
\centering
\includegraphics[clip,width=250pt]{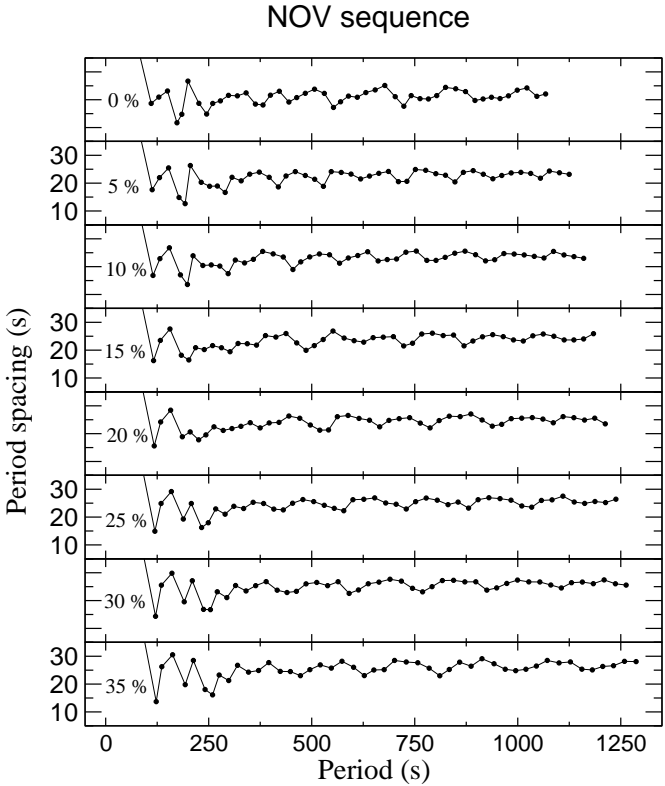}
\caption{The forward period spacing in terms of the periods for $\ell=
2$  modes corresponding  to the  NOV model-sequence  analyzed  in Fig.
~\ref{pcrnov}.  Each   panel  shows  the   predictions  for  different
percentages  of   the  crystallized  mass   fraction.   Here  chemical
rehomogenization induced by crystallization has been considered.}
\label{dpcrnov}
\end{figure}

\begin{figure}
\centering
\includegraphics[clip,width=250pt]{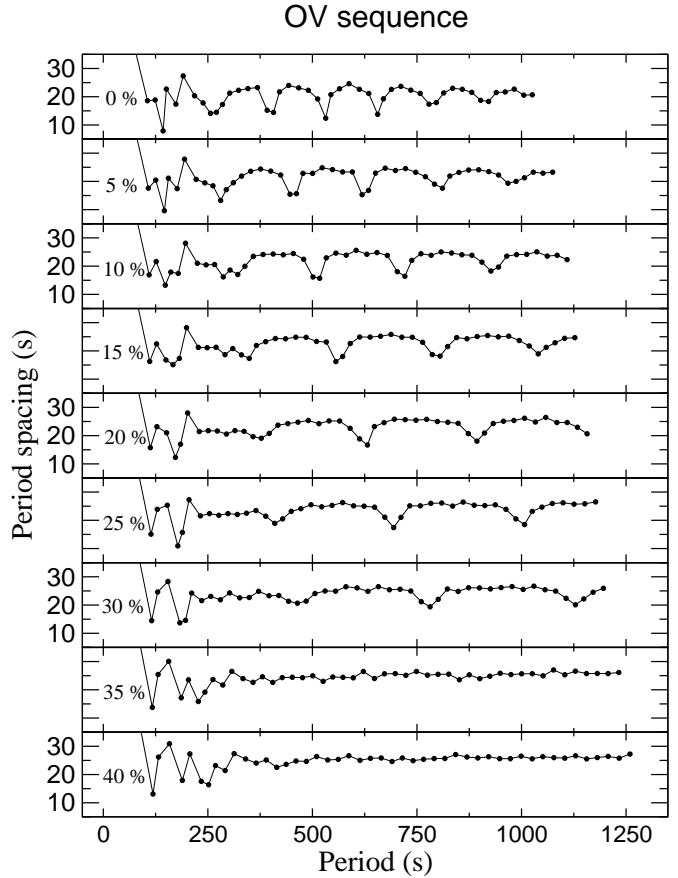}
\caption{Same as Fig. ~\ref{dpcrnov}, but for the OV model-sequence.} 
\label{dpcrov}
\end{figure}

\subsection{The Brunt-V\"ais\"al\"a frequency}

The procedure we  follow to assess the run  of the Brunt-V\"ais\"al\"a
frequency ($N$)  is that proposed  by Brassard et al.   (1991b).  This
numerical treatment  accounts explicitly  for the contribution  to $N$
from any change  in composition in the white dwarf  models by means of
the Ledoux term  $B$.  This is an important  aspect in connection with
the  phenomenon   of  mode  trapping  and   confinement  (Brassard  et
al.  1992a;  C\'orsico   et  al.  2002).   In  the   lower  panels  of
Figs. ~\ref{xbvnov} and ~\ref{xbvov} we show the run of $N^2$ in terms
of  the outer mass  fraction for  white dwarf  models at  $T_{\rm eff}
\approx 11800$ K corresponding  to sequences NOV and OV, respectively.
Element diffusion strongly smooths  out the external chemical profile
emerging from the  thermally pulsing AGB phase to  such an extent that
the Brunt-V\"ais\"al\"a  frequency in  the outer layers  exhibits very
smooth  local features.  Note  that because  of the  markedly distinct
shape  of the  chemical distribution  in the  innermost region  of the
model  with core  overshooting,  the  profile of  $N$  for such  model
exhibits   a  clearly   peaked   feature  at   $\log   q  \sim   -0.7$
(Fig. ~\ref{xbvov}).  This feature  is responsible for the presence of
pronounced minima in the period spacing distribution (see figure 15 of
Paper  I).  The  model corresponding  to  sequence NOV  also shows  an
innermost  peak in $N$  (Fig. ~\ref{xbvnov}).  However, because  it is
located at a  deeper layer and has a smaller amplitude  than in the OV
sequence, it produces a much smaller amount of mode trapping.

\subsection{Results without crystallization}

In this section we shall describe pulsational results in which we have
neglected  crystallization  in  the  computation  of  the  pulsational
spectrum.   In Fig.   ~\ref{pncnrnov}  we show  the  evolution of  the
$\ell= 2$ periods in terms  of $T_{\rm eff}$\ corresponding to the NOV
sequence of models.  Clearly,  the periods increase monotonically with
decreasing $T_{\rm eff}$, particularly in  the case of modes with high
radial  order $k$.  This  effect is  explained on  the basis  that the
Brunt-V\"ais\"al\"a  frequency at the  core decreases  as a  result of
increasing degeneracy in the core.  Note  that the mode with $k= 1$ is
rather  insensitive to  the  white dwarf  cooling,  though its  period
slightly grows with decreasing $T_{\rm eff}$.  We note that the curves
are very smooth in all the $T_{\rm eff}$ range considered, and only in
a few  cases is  there a slight  approach between periods  of adjacent
modes.  In Fig.  ~\ref{pncnrov} we  depict the evolution of periods in
the  case of models  corresponding to  OV sequence.  In this  case the
overall trend of  the periods is similar to that  found in NOV models,
but  they show  instead  clear  signals of  mode  bumping and  avoided
crossing\footnote{When a  pair of modes  experiences avoided crossing,
the modes exchange their  intrinsic properties after mode bumping; see
Aizenman et al.  (1977).}.  For  instance, a series of mode bumping is
observed between modes with $k= 17$  and $k= 18$, $k= 18$ and $k= 19$,
$k=  19$  and $k=  20$,  and  $k= 20$  and  $k=  21$  (thick lines  in
Fig. ~\ref{pncnrov}).  We also find  that, apart from the periods, the
kinetic energy  and period spacing  minima are also  exchanged between
these  pairs of modes  after mode  bumping has  taken place.   A close
inspection  of the  eigenfunctions and  weight  functions demonstrates
that modes experiencing avoided crossings are {\sl core trapped} ones,
that is, modes with relatively  large amplitudes in the region bounded
by the centre of the model and  the location of the step in the oxygen
profile   left  by   overshooting   (see  Paper   I).   Similar   mode
bumping/avoided crossing  features in white  dwarf evolutionary models
have  been reported  by  Wood \&  Winget  (1988) and  Brassard et  al.
(1991a),  although  in such  calculations  modes experiencing  avoided
crossings  were identified  as  modes trapped  in  the outer  hydrogen
envelope.  In the context of massive white dwarf models, Montgomery \&
Winget  (1999)   have  found  avoided  crossing   phenomena  when  the
crystallized mass fraction  of their models is varied  in a continuous
fashion and  the other  model parameters (such  as $T_{\rm  eff}$) are
held fixed.

\begin{figure*}
\centering
\includegraphics[width=350pt]{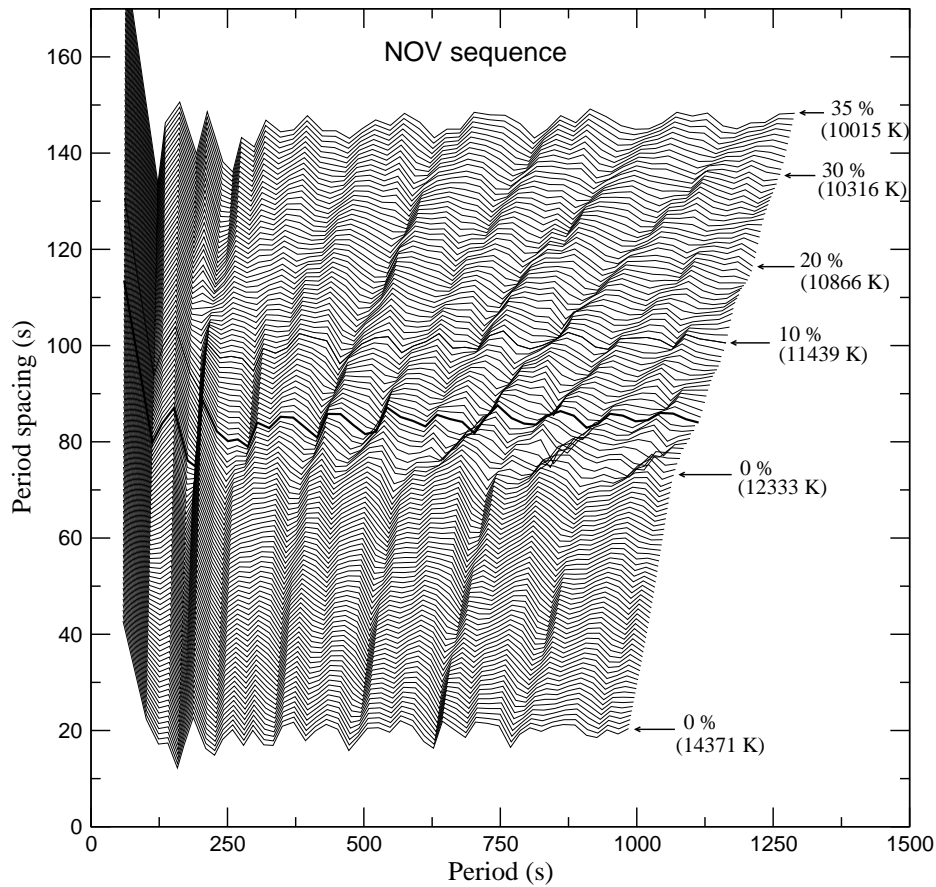}
\caption{Period  spacing vs.  period for  effective temperature values 
between  $\approx$  15000  and   10000  K  corresponding  to  the  NOV
model-sequence.  In the  interests  of clarity,  each  curve has  been
shifted  upward starting  from the  lowest  one, in  increments of  1
sec. For details, see the text.}
\label{dpfullnov}
\end{figure*}

\begin{figure*}
\centering
\includegraphics[width=350pt]{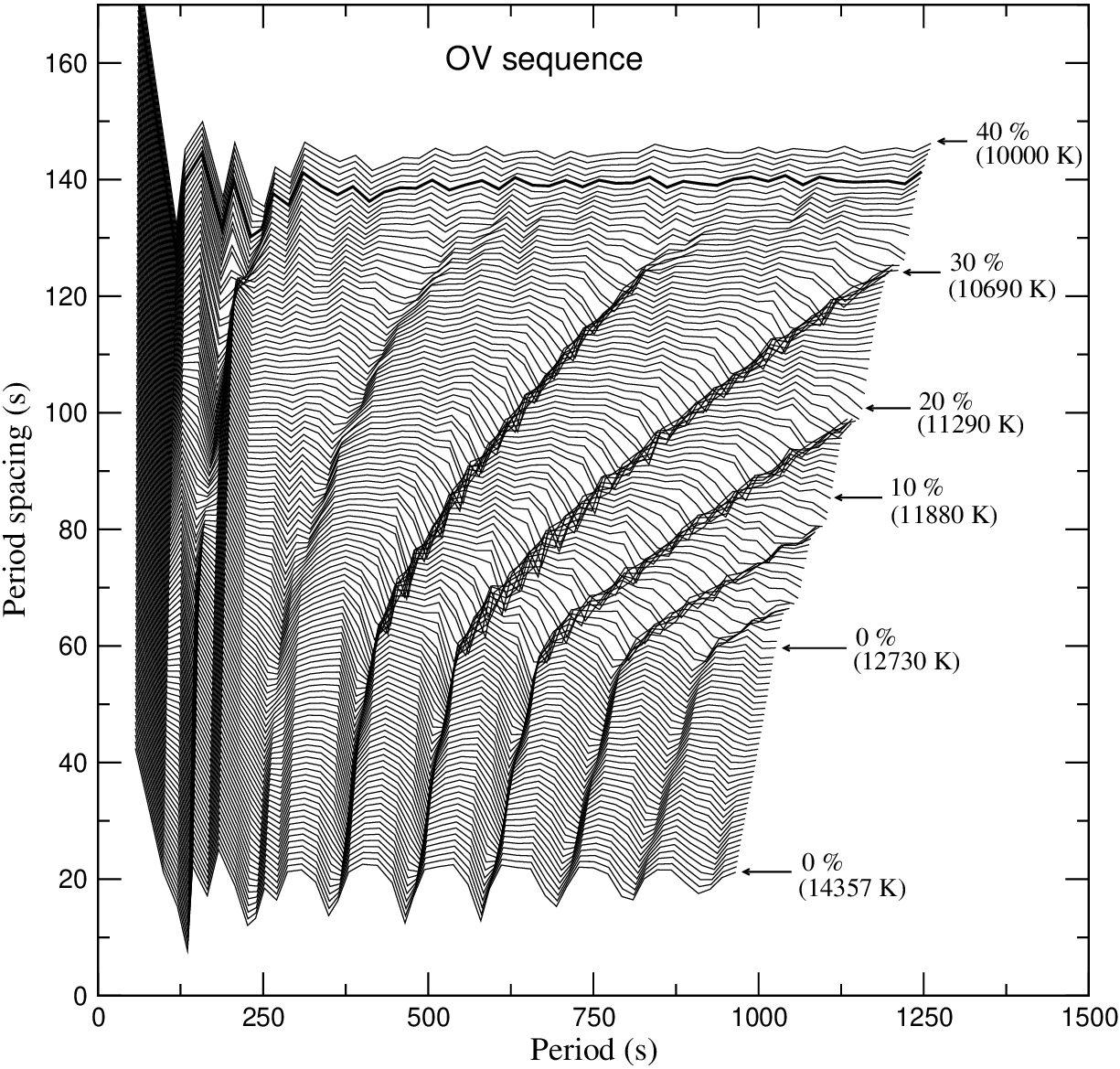}
\caption{Same as Fig. ~\ref{dpfullnov}, but for the OV model-sequence.} 
\label{dpfullov}
\end{figure*}

\subsection{Results considering crystallization without chemical 
rehomogenization}

In this section,  we describe our results when account  is made of the
presence  of  a solid  core  in  the  computation of  the  pulsational
eigenspectra.   However,  we  neglect  here  the  effect  of  chemical
rehomogenization due  to phase separation, that is,  the core chemical
profile remains fixed as the white dwarf crystallizes.  We stress that
we take the crystallized mass fraction  not as a free parameter but as
given   by  evolutionary   computations.   Figs.    ~\ref{cnrnov}  and
~\ref{cnrov}  depict the resulting  evolution of  periods in  terms of
$T_{\rm eff}$  for sequences NOV  and OV, respectively.  Note  that in
both  cases   the  onset  of  crystallization  is   reflected  in  the
pulsational spectrum as a rather  abrupt increase in periods for modes
with high- and intermediate radial orders, whereas lower overtones are
rather insensitive to  the presence of the solid  core.  This fact can
be  explained  on  the  basis  that  for $g$-modes,  when  $k  \gg  1$
(asymptotic  limit)   the  periods  are  given  by   $P_k  \propto  k\
(\int^{r_2}_{r_1} N dr / r)^{-1}$  (Tassoul et al.  1990), where $r_1$
and   $r_2$   define   the   propagation  region   of   modes.    When
crystallization is considered,  the internal boundary $r_1(M_{\rm c})$
moves outward, so the integral  $\int^{r_2}_{r_1} N dr / r$ decreases,
and consequently  the periods (and  the period spacings)  increase. An
important feature  displayed by Figs.   ~\ref{cnrnov} and ~\ref{cnrov}
is  that,  when   crystallization  begins,  the  mode  bumping/avoided
crossing   phenomena  propagate  to   longer  periods.    Indeed,  the
outward-moving  crystallization front reinforces  the process  of mode
bumping/avoided crossing,  which are also present before  the onset of
crystallization.  This  effect is particularly noticeable  in the case
of OV  models.  In Figs.   ~\ref{dpcnrnov} and ~\ref{dpcnrov}  we show
the period  spacing diagrams  corresponding to various  percentages of
the  crystallized  mass  fraction  for  NOV and  OV  model  sequences,
respectively. Note that in the case  of OV models the strong minima in
the  period  spacing are  gradually  displaced  to  longer periods  as
crystallization proceeds.  At  the same time they move  away from each
other.   Note that,  at $T_{\rm  eff} \approx  10000$ K  ($ 40  \%$ of
crystallized mass fraction)  only one minimum at about  550 s remains.
The period spacing diagrams exhibited by models with core overshooting
are completely different from those of models in which overshooting is
neglected,  as can be  inferred from  Fig. ~\ref{dpcnrnov}.   In fact,
although  the  period  spacing   diagrams  are  strongly  modified  by
crystallization, their overall  structure looks very different between
both sequences.   In particular, sequence NOV lacks  the strong minima
characterizing sequence  OV, features which, as concluded  in Paper I,
could be  eventually used  to place constraints  on the  occurrence of
core overshooting.   {\sl So,  the conclusions arrived  at in  Paper I
concerning  the role  of core  overshooting in  massive ZZ  Ceti stars
remain valid even when the effect of a solid core on the pulsations is
included}.  However,  this comes with the  caveat that if  the star is
more  massive than  this  model, then  it  could have  a large  enough
crystallized mass  fraction ($\sim 80\,$\%) so that  the chemical step
left  by overshooting  would be  engulfed by  the  crystallizing core,
erasing this  feature.  In  this case we  should expect  a pulsational
pattern with no  appreciable minima in the period  spacing. As we will
see  in the  following  section,  this situation  can  be reached  for
smaller amounts of crystallization  if the effects of phase separation
and mixing are taken into account.

\subsection{Results considering crystallization and chemical 
rehomogenization}

As  recently  shown,  the  occurrence of  core  overshooting  strongly
modifies the appearance of the period diagrams even in the presence of
a solid  core.  Now, theoretical  evidence strongly suggests  that, if
the white dwarf core is composed  initially of a mixture of carbon and
oxygen, then  the crystallized region will have  an enhanced abundance
of oxygen  compared to that in  the original fluid  state (Ichimaru et
al.  1988; Segretain  \& Chabrier 1993, Salaris et  al. 1997).  On the
other   hand,   the   fluid   regions   overlying   the   crystallized
oxygen-enhanced  layers will have  a higher  content of  carbon.  This
region will become Rayleigh-Taylor  unstable, since carbon is slightly
less dense  than oxygen. This instability leads  to a rehomogenization
of the  chemical profile,  and, as a  result, the  chemical abundances
after crystallization can differ  substantially when compared with the
initial ones.  This is an  important point because any feature present
in  the  chemical profile  could  be  potentially  wiped out  by  this
rehomogenization process,  even in fluid layers located  far away from
the crystallization front.

\begin{figure}[t]
\centering
\includegraphics[width=\columnwidth]{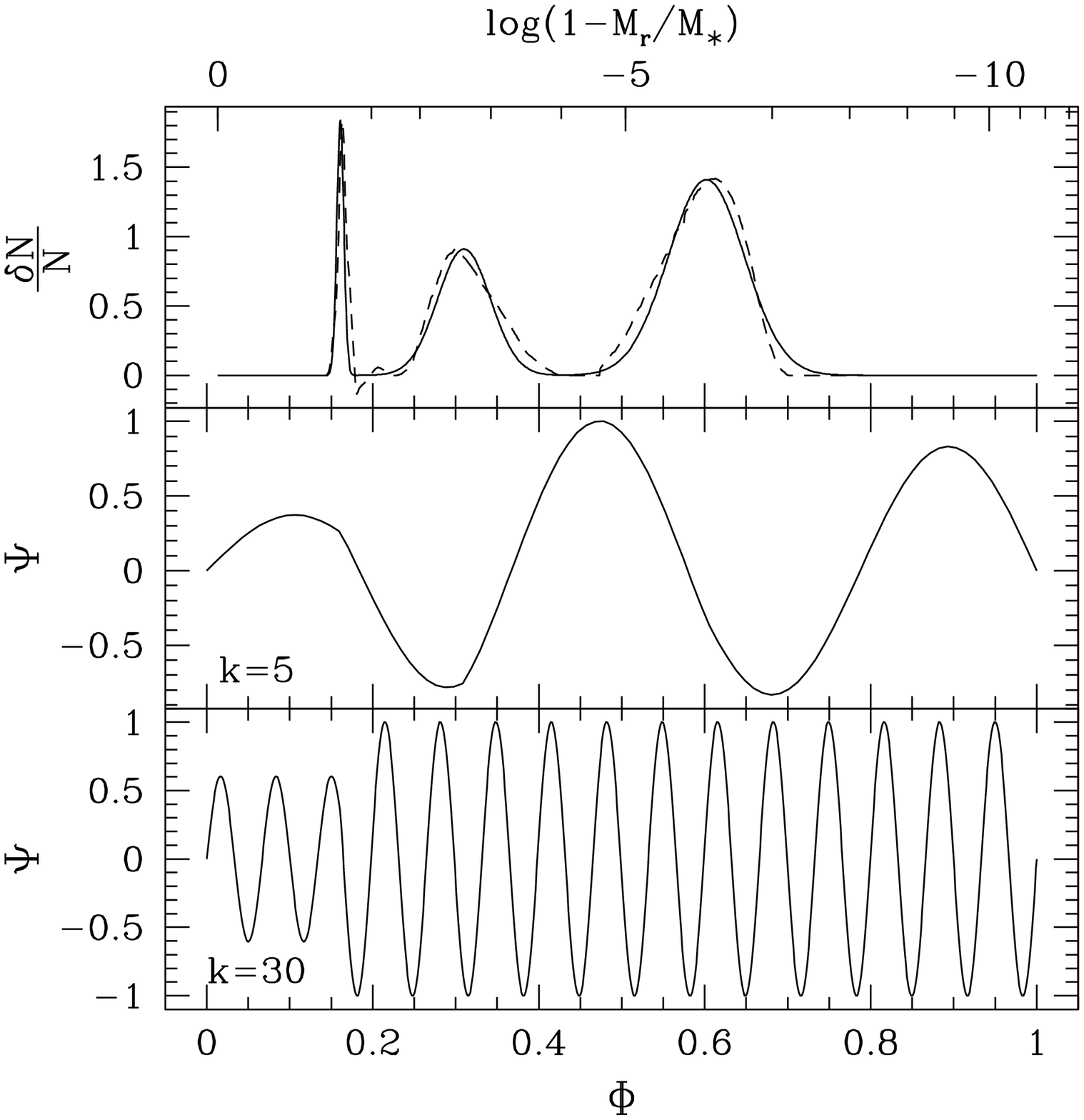}
\caption{Top  panel: the  bumps/non-smooth  features  in the  \bvfreq\
found in  our numerical  models (dashed curve)  and the  assumed bumps
which we  use in  our simple analytical  model (solid curve).   In the
lower panels, we illustrate how  these features can act to trap modes.
For instance,  the narrowness of the $\Phi=0.16$  feature, compared to
the spatial wavelength  of these modes, causes it  to induce amplitude
changes in the two eigenmodes pictured.  }
\label{pert}
\end{figure}

\begin{figure}[t]
\centering
\includegraphics[width=\columnwidth]{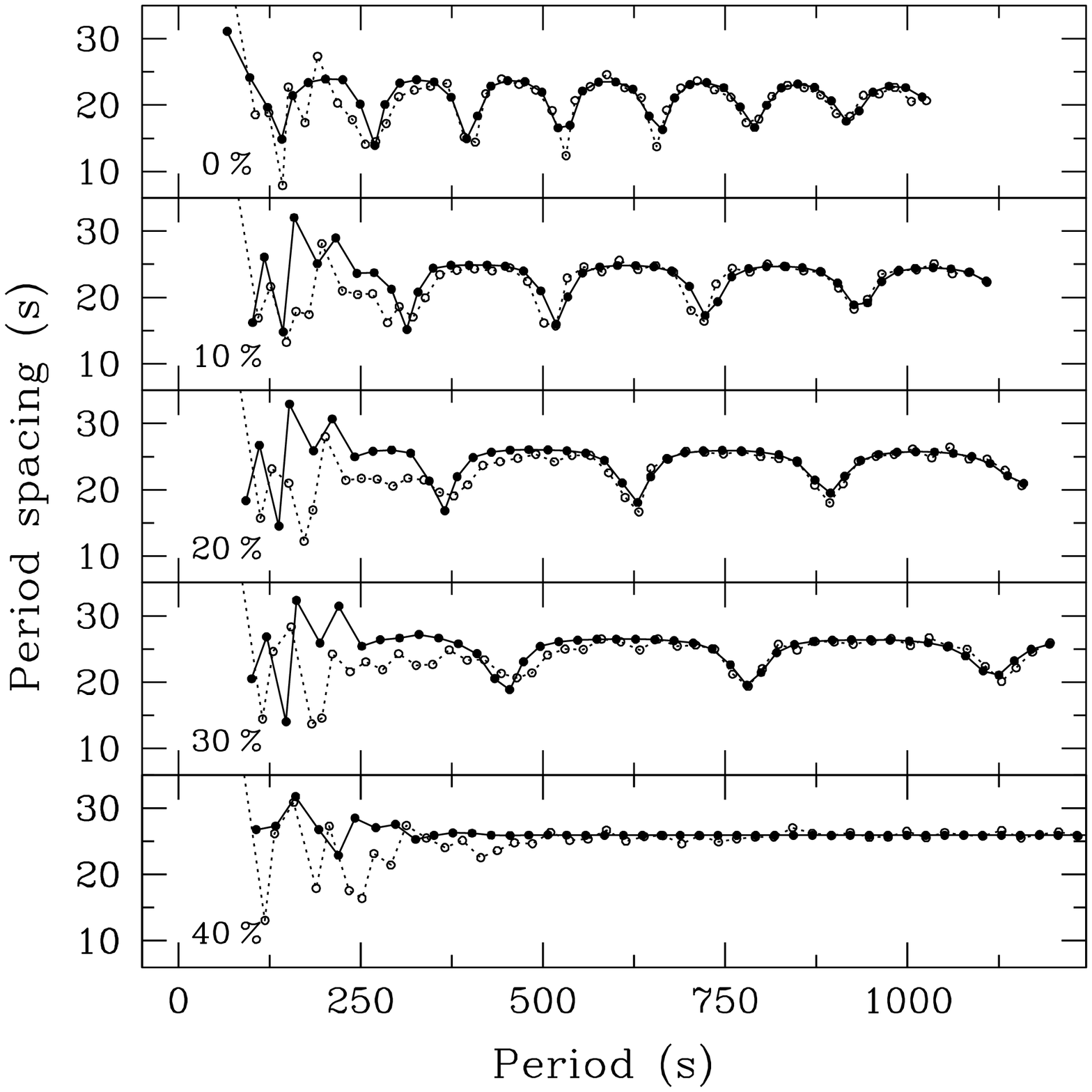}
\caption{Comparison  of the  mode trapping  of  the  analytical  model 
(filled circles,  solid lines) to  that of the full  numerical results
shown in Fig.~13 (open circles, dotted lines).  Each panel is labelled
with the crystallized mass fraction to which it corresponds. }
\label{mtrap}
\end{figure}

To  compute  the above-mentioned  chemical  rehomogenization, we  have
employed  the same  algorithm  as  in Montgomery  et  al.\ (1999)  and
Salaris et~al.\ (1997). To derive  the oxygen enhancement when a given
layer  crystallizes,  we  adopt   the  Segretain  \&  Chabrier  (1993)
spindle-type phase diagram for  a carbon/oxygen mixture. In our scheme
for mixing,  we first  consider a crystallized,  oxygen-enhanced layer
and  then we  ask if  the innermost  fluid shell  has a  higher carbon
content than  the overlying one.  If it does,  then we mix  both fluid
layers and  perform the  same comparison with  the next  layer further
out. In  this way we  scan outward through  the fluid and  the process
stops when  further mixing no  longer decreases the carbon  content of
the fluid  between this point  and the crystallization  boundary. When
the  crystallization   front  moves   outward  due  to   cooling,  the
aforementioned procedure  is repeated.  In Fig.  ~\ref{rehomo} we show
the oxygen chemical profile in terms of the fractional mass for the OV
model  sequence corresponding to  various degrees  of crystallization.
The  initial, uncrystallized  profile is  depicted with  a thick-solid
line, and the resulting  profiles after the rehomogenization has taken
place  are  shown  with  thin  lines  for  10,  20,  30  and  40$\,$\%
crystallized  mass fractions.  Note  that the  shape  of the  chemical
profile  is   strongly  modified,  even   in  regions  far   from  the
crystallization  front.  As  a clear  example  of this,  we show  with
thick-dashed line the oxygen profile at the moment that the pronounced
feature  at $M_r  \approx 0.80  M_*$  (left by  core overshooting)  is
completely  eliminated.   This  occurs   when  the  location   of  the
crystallization boundary reaches $M_{\rm c} \approx 0.37 M_*$.

We wish now to  investigate how such chemical rehomogenization affects
the   pulsational  spectrum   of   the  white   dwarf  models.   Figs.
~\ref{pcrnov}  and ~\ref{pcrov}  document the  evolution of  $\ell= 2$
periods in  terms of $T_{\rm eff}$  corresponding to cases  NOV and OV
sequences,  respectively.   In  the  case  of the  NOV  sequence,  the
rehomogenization process  induced by crystallization  clearly modifies
the internal chemical profile (it  smooths out the profile at $\log q
\approx   -0.3$)  to   such  a   degree  that   any  signal   of  mode
bumping/avoided  crossing  is  almost  absent, in  contrast  with  the
situation  in   which  such  mixing  has  been   neglected  (see  Fig.
~\ref{cnrnov}). In the case of  OV models, the theoretical spectrum of
periods barely  changes in  response to the  rehomogenization (compare
Figs.   ~\ref{cnrov} and  ~\ref{pcrov}).  That  is, clear  features of
mode bumping/avoided crossing survive  even during the stages in which
the pronounced  step in the oxygen profile  is continuously diminished
by  chemical rehomogenization (see  Fig.  ~\ref{rehomo}).   However, a
closer  inspection of  Fig.   ~\ref{pcrov} reveals  that from  $M_{\rm
c}/M_* \approx  0.37$ ($T_{\rm eff}  \approx 10400$ K)  these features
virtually   disappear,   and  the   period   spacing  becomes   almost
uniform.  This  effect  can  be  appreciated more  clearly  from  Fig.
~\ref{dpcrov}, in  which we have  plotted period spacing  diagrams for
several degrees  of crystallization.  Note that  the pronounced minima
characterizing the  period spacing distribution at  periods $\sim 760$
and $\sim 1120$  s (for $30\,\%$ crystallization) are  absent when the
crystallized mass  fraction is  $35 \%$.  In  fact, when  the chemical
discontinuity   has   been    completely   wiped   out   by   chemical
rehomogenization,  the predicted  period spacing  distribution becomes
very flat, particularly for periods longer than $\sim 300$ sec.

Figs.   ~\ref{dpcrnov} and  ~\ref{dpcrov} represent  some of  the main
results of  the present  work. A comparison  of these  figures clearly
indicates that  the period spacing distribution for  sequences NOV and
OV is markedly different.  In particular,  we note that in the case of
the OV model  with $M_{\rm c}= 0.35 M_*$,  after the overshoot-induced
step has  been almost  removed from the  chemical profile,  the period
spacing distribution is notably more  uniform as compared with the NOV
model characterized by the same percentage of crystallization. This is
because in regions  farther out from the centre ($\log  q \la -1$) the
OV model is actually \emph{smoother} than the NOV model, so its period 
spectrum shows less mode trapping.

Finally,  in Figs.  ~\ref{dpfullnov} and  ~\ref{dpfullov} we  show the
forward  period spacing  vs period  for 15000  K $\gtrsim  T_{\rm eff}
\gtrsim  10000$ K  corresponding to  the NOV  and  OV model-sequences,
respectively.  For  the sake of  clarity, each curve has  been shifted
upward starting from the lowest  one, in increments of 1 sec.  Labels
of the solid curves indicate the percentage of crystallization and the
corresponding  $T_{\rm   eff}$  of   the  models.  The   dashed  curve
corresponds  to the  moment at  which  the relevant  structure in  the
oxygen  profile (at  $\log q  \approx  -0.3$ in  the case  of the  NOV
sequence, and at $\log q \approx -0.7$ in the case of the OV sequence)
disappears due to chemical  rehomogenization. From these plots one can
see  the evolution of  the trapped  modes quite  clearly, particularly
when  crystallization  begins.  Note  from  Fig.  ~\ref{dpfullov}  the
decrease   in  amplitude   of   mode  trapping   as   the  degree   of
crystallization  increases. This  figure also  emphasizes  the gradual
change experienced by the period spacing when the step  left by
overshooting is removed by chemical rehomogenization.

\section{Analytical model of mode trapping}
 
In  this  section  we examine  the  mode  trapping  in our  models  by
comparing  it to  the  results  of a  simple  analytical model.   This
analytical model,  which will be  described in more  detail elsewhere,
describes  the  high-overtone  limit  of the  pulsation  equations  as
derived  by Deubner  \&  Gough  (1984) and  Gough  (1993).  While
formally equivalent to the  standard linear adiabatic wave equation in
the Cowling approximation  --- see, e.g., Cox (1980)  --- this form of
the   equations  reduces   the  problem   to  a   single  second-order
differential  equation resembling  the oscillations  of  a non-uniform
string.  In fact, using this  equation as a convenient starting point,
Montgomery, Metcalfe, \& Winget (2003) showed that a symmetry inherent
in the  string problem  is still present  in stellar  pulsations. This
lead to the discovery of the core/envelope symmetry and the subsequent
realization that the chemical composition gradients in the core may be
even  more effective  in producing  mode  trapping than  those in  the
envelope.  Here we use an extended version of this model in which the
perturbations to the background state  need not be small, although our
solutions are still formally valid only in the high-overtone limit.

In Fig.~\ref{pert},  we illustrate how features/bumps  in the \bvfreq\
result  in  mode trapping  in  the  eigenfunctions.  As in  Montgomery
et~al.\  (2003), we  use the  ``normalized buoyancy  radius'', $\Phi$,
defined by 
\[
\Phi(r) \equiv \frac{\int_{0}^{r} dr\,\frac{|N|}{r}}{
    \int_{0}^{R_{\star}} dr\,\frac{|N|}{r}},
\]
as the radial  variable, since with this choice  the wavelength of the
eigenfunctions appears to be  constant, making the correspondence with
the vibrating string  more apparent ($\Phi$ equals 0  and 1 correspond
to the centre  and surface of the model,  respectively).  In addition,
the    eigenfunction    $\Psi    \approx   (r^2/g    \rho^{\mbox{\tiny
1/2}})\,\delta  p$ is  defined  in such  a  way that  it has  constant
amplitude except in regions where the background state varies rapidly,
such as the  bumps shown in the top panel  of Fig.~\ref{pert}.  As can
be  seen,  the ``bump''  at  $\Phi \sim  0.16$  is  narrower than  the
wavelengths of either the $k=5$ or $k=30$ mode, so both modes perceive
it  essentially   as  a   delta  function  and   therefore  experience
significant  mode trapping (amplitude  change) in  propagating through
it.   In contrast,  the other  two ``bumps''  at $\Phi  \sim  0.3$ and
$\Phi\sim 0.6$ are much wider. Given that the wavelength of the $k=30$
mode  is  smaller  than  the  width of  these  bumps,  it  experiences
essentially no mode trapping. On the other hand, the wavelength of the
$k=5$ mode  is somewhat larger  than the width  of these bumps,  so it
still experiences a modest  amount of mode trapping (amplitude change)
due to these features.

In the top panel of Fig.~\ref{pert}  we have chosen the height, width,
and position of the bumps in the analytical model (the solid curve) to
mimic those found in a full evolutionary model (the dashed curve).  In
Fig.~\ref{mtrap} we  show  the mode  trapping found in  our analytical
model (filled  circles, solid lines) and  that  found in the  OV model
shown  in Fig.~13  (open  circles, dotted lines),  for  values  of the
crystallized mass fraction between  $0\,$\% and $40\,$\%. Clearly, the
overall  correspondence is quite good,  except for  low periods, where
our asymptotic model  is  anyway not  applicable.  In particular,  the
asymptotic model  is able to reproduce   i) the shape of  the trapping
cycle (sharp minima and flat maxima), ii) the decrease in amplitude of
mode  trapping for high  periods/overtones,  iii) the increase in  the
period of  the   trapping cycle  and   the movement of  mode  trapping
features to  higher  periods  as  the  degree   of  crystallization is
increased,  and iv) the absence  of trapping features for the $40\,$\%
crystallized case. 

The  explanation  of  these  effects is  straightforward.  First,  the
overall good  match is due to  the fact that the  modes shown actually
have  high  overtone numbers  \emph{and}  that  the  mode trapping  is
dominated by the  single feature in $\delta N/N$  at $\Phi \sim 0.16$.
Second, the decrease in trapping with increasing period occurs because
these  modes   have  shorter  spatial  wavelengths,   and  when  these
wavelengths become comparable to the  width of the features in $\delta
N/N$  the modes  cease to  feel the  effects of  a  sudden transition.
Third, the  reason that  the number of  modes per mode  trapping cycle
increases is that crystallization effectively moves the $\Phi=0$ point
to the right, making the bump  at $\Phi \sim 0.16$ closer to the inner
turning  point. The  shortest possible  trapping cycle  of 2  modes is
achieved for a  feature in the middle at $\Phi \sim  0.5$, so moving a
bump closer  to an edge  has the effect  of making the  trapping cycle
longer.  Finally,   the  lack  of   mode  trapping  in   the  $40\,$\%
crystallized case is  due to the removal of the  $\Phi \sim 0.16$ bump
due   to  crystallization,  phase   separation,  and   the  subsequent
rehomogenization:  the   remaining  bumps  cause   virtually  no  mode
trapping.

As a final note we point out that the core/envelope symmetry found by
Montgomery et~al.\ (2003) is not an issue for these models, at least
for modes with periods greater than 300 sec. This is because,  as
  is conclusively demonstrated in Fig.~17, the outer chemical
transition zones (C/He and He/H) are found to be so smooth that they
produce almost no mode trapping; thus, any mode trapping \emph{must}
be a result of structure in the deeper regions.

\section{Discussion and conclusions}

Convective overshooting  is a  longstanding problem  in the  theory of
stellar  structure and evolution.  It  is   well known on  theoretical
grounds  that  during many stages  in   their lives  stars  experience
overshoot episodes, that   is,  partial  mixing beyond  the   formally
convective boundaries  as predicted by  the Schwarzschild criterion of
convective stability (Zahn  1991; Canuto 1992; Freytag   et al.  1996;
see also Renzini 1987).  In particular, core overshooting taking place
during  central burning   is   an  important   issue because    it has
significant  effects on the  stellar structure and evolution. Over the
years,   considerable  observational effort    has been    devoted  to
demonstrating  the   occurrence   of    core overshooting.     Indeed,
confrontation of  stellar models with a  wide variety of observational
data suggests that convective overshoot  takes place in real stars ---
see Stothers \& Chin (1992),  Alongi et al. (1993),  Kozhurina-Platais
et al.  (1997),  Herwig et al. (1997), von  Hippel \& Gilmore  (2000),
among others. 

However,  most   of  the  evidence   about  the  occurrence   of  core
overshooting  relies primarily  on  observational data  from the  very
outer layers of stars from  where radiation emerges.  A more promising
and  direct  way of  placing  constraints  on  the physical  processes
occurring in the very deep interior  of stars is by means of the study
of   their  pulsational  properties.    Pulsating  white   dwarfs  are
particularly  important  in  this   regard.   In  fact,  white  dwarfs
constitute the end product of  stellar evolution for the vast majority
of  stars,  and  the  study  of  their  oscillation  spectrum  through
asteroseismological techniques has become  a powerful tool for probing
the  otherwise  inaccessible inner  regions  of  these stars  (Bradley
1998b;   Metcalfe  et   al.   2002;  Metcalfe   2003).   White   dwarf
asteroseismology has  also opened the  door to peer into  the physical
processes that lead to the formation of these stars.

In this  work ---  and also  in Paper I  --- we  have argued  that the
occurrence  of  core overshoot  episodes  during  core helium  burning
leaves strong  imprints on the theoretical period  spectrum of massive
ZZ Ceti stars, features which could in principle be used for providing
strong constraints  on the occurrence  of such episodes. On  the other
hand, we have demonstrated  that the chemical rehomogenization induced
by phase separation  gives rise to a featureless  period spectrum when
the crystallized mass fraction is  larger than $\approx 30 \%$. One of
the aims of the present paper  was to place constraints on the stellar
mass and \teff\ values at which we should expect a pulsational pattern
without any  signature of  core overshoot.  We  have seen that  in our
sequence OV  the chemical discontinuity  left by core  overshooting is
wiped out  by chemical rehomogenization  for a solid core  larger than
$37 \%$, which  for a $0.94$-\msun white dwarf  takes place at $T_{\rm
eff}=  10400$ K.  The effective  temperature at  which this  occurs is
strongly  dependent on the  stellar mass.  For instance,  for $1.00$-,
$1.03$-,  and $1.05$-\msun  white  dwarf models,  the  traces of  core
overshooting  are  removed  by  chemical rehomogenization  at  $T_{\rm
eff}=$ 11700, 12350, and 12800~K, respectively.

The white  dwarf star  BPM 37093,  the most massive  ZZ Ceti  known to
date, is  particularly noteworthy in  this regard. Using  IUE spectra,
Koester \&  Allard (2000) derive  \teff$=11520$ K, $\log g=  8.67$ and
$M_*=  1.03$ \msun.   Other recent  determinations place  the  mass of
BPM~37093 at $1.00\,\msun $ (Bergeron et al.  2001) and $1.10\,\msun $
(Bergeron  et al.\  2004; Fontaine  et al.\  2003).  Even  given these
uncertainties,  it  seems likely  that  its  mass  lies in  the  range
1.0--1.1$\,\msun $. With regard to its pulsation properties, BPM 37093
exhibits  $g$-mode pulsations with  periods in  the range  $500-660$ s
(Kanaan et al.  2000). The periods are very accurately determined, but
as  far  as  we  are   aware,  the  $\ell$  values  are  not  securely
identified\footnote{By applying the $\ell$-value identification method
of  Robinson et al.  (1995), Nitta  et al.  (2000) have  discarded the
possibility that the main periods  observed in BPM 37093 are $\ell= 3$
modes, but they cannot  give definite identifications regarding $\ell=
1$ or $\ell= 2$ values.}.  On  general grounds, it is possible to show
that  the observed  modes  cannot  all be  $\ell=1$,  so the  simplest
assumption has been that they  are $\ell= 2$ modes\footnote{If all the
modes are $\ell= 1$ then the average period spacing would imply a mass
much   larger   than  the   values   found   from  the   spectroscopic
determinations.}   (Nitta  et  al.  2000, Kanaan  et  al.   2000,
Montgomery \& Winget 1999).  Under this assumption, it is possible to
construct   an   ``observed''  period   spacing   diagram,  which   is
characterized  by  an  average  period  spacing  of  $17.125$  s  with
deviations due to mode trapping of $\approx \pm 8$~s.

Given the uncertainties  in the mass and \teff\ of  BPM 37093 and also
in  its  mode identification,  we  cannot  make  a definite  statement
concerning either whether core overshooting took place in BPM 37093 or
what degree  of crystallization it  possesses.  However, we  can place
constraints  on   the  different  possibilities.   In   fact,  we  can
characterize BPM 37093 by a stellar mass in the range $1.00\ \msun \la
M_*  \la 1.10$  \msun\ and  an effective  temperature in  the interval
$11000 \la T_{\rm  eff} \la 12000$ K, given  the present uncertainties
in the derivation of these quantities.  In the case of 1.00 \msun\ our
treatment  of  crystallization  predicts  that  the  model  should  be
$37\,\%$  crystallized  at  $\approx  11700$ K.   Thus,  its  observed
frequency spectrum  could reflect mode  trapping effects due  to prior
evolutionary  episodes  of core  overshoot.   On  the  other hand,  if
BPM~37093 has a mass of $M_* \approx 1.10$ \msun, then there should be
virtually no mode  trapping features in its period  spectrum.  In this
case, we  cannot say anything  about the occurrence of  core overshoot
during  pre-white  dwarf evolution.   A  possible  explanation of  the
observed period spectrum would be  that the observed modes are not all
$\ell=2$, $m$=0 modes, but instead  a mixture of $\ell=1$ and $\ell=2$
modes  (Metcalfe et al. 2004, Nitta et al. 2000).

\section{Concluding remarks} 

This  paper is  aimed  at specifically  assessing  the feasibility  of
employing white  dwarf asteroseismology to  demonstrate the occurrence
of core overshooting during core helium burning.  The investigation is
focused on massive, intermediate-mass stars, the complete evolution of
which has  been followed from  the zero-age main sequence  through the
thermally pulsing  phase on the  asymptotic giant branch (AGB)  to the
white dwarf  regime.  This allows us  to obtain ZZ Ceti  models with a
physically sound internal structure consistent with the predictions of
stellar evolution theory. Our work shows that for studying the effects
of  core overshooting,  pulsating white  dwarfs with  masses  of $\sim
0.90\,\msun$ would  be ideal  since the potential  for crystallization
will not  be a factor in  erasing the signature  of these overshooting
events.   Conversely,  for studying  the  effects of  crystallization,
masses $\ga 1.0\,\msun$ are best since most of the core structure will
be erased  by crystallization and the chemical  mixing associated with
phase separation, making it easier  to isolate and study the effect of
crystallization on its own. In  future studies we will focus on models
in these mass ranges.

\begin{acknowledgements}

We warmly acknowledge to our referee Donald Winget for his suggestions
and comments  that improved  the original version  of this  work. This
research was supported by  the Instituto de Astrof\'{\i}sica La Plata,
by  the UK  Particle Physics  and Astronomy  Research Council,  by the
CIRIT, by MCYT grants AYA2002-04094-C03-01 and 02, and by the European
Union  FEDER funds.   LGA also  acknowledges  the Spanish  MCYT for  a
Ram\'on y Cajal fellowship.

\end{acknowledgements}

\end{document}